\documentclass[floatfix,superscriptaddress,showpacs,amssymb,10pt,aps,prd,reprint,longbibliography]{revtex4-1}

\usepackage{graphicx,epsfig,amssymb} 
\usepackage{amsmath,amsfonts, times}
\usepackage{bm} 

\usepackage[linktocpage,colorlinks]{hyperref}
\usepackage[caption=false]{subfig}
\usepackage[usenames]{color}     
\usepackage{natbib}
\usepackage{soul}
\usepackage[utf8x]{inputenc}
\usepackage{float}

\definecolor{coolblack}{rgb}{0.0, 0.18, 0.39}
\definecolor{darkred}{rgb}{0.5,0,0}
\definecolor{darkgreen}{rgb}{0,0.5,0}
\definecolor{darkblue}{rgb}{0,0,0.5}
\definecolor{lapislazuli}{rgb}{0.15, 0.38, 0.61}
\definecolor{venetianred}{rgb}{0.78, 0.03, 0.08}
\definecolor{bleudefrance}{rgb}{0.19, 0.55, 0.91}
\definecolor{dogwoodrose}{rgb}{0.84, 0.09, 0.41}
\hypersetup{colorlinks=true, citecolor=darkgreen, linkcolor=darkblue, 
	urlcolor = blue}

\newcommand{\dd}{\mathrm{d}}
\newcommand{\expe}{\operatorname{e}}

\begin{document}
	
	\author{Qian Li}
	\address{School of Physics and Technology, Wuhan University, Wuhan, 430072, China}
	\author{Qianchuan Wang}
	\address{School of Physics and Technology, Wuhan University, Wuhan, 430072, China}
	
	\author{Junji Jia}
	\email[Corresponding author:~]{junjijia@whu.edu.cn}
	\address{Department of Astronomy $\&$ MOE Key Laboratory of Artificial Micro- and Nano-structures, School of Physics and Technology, Wuhan University, Wuhan, 430072, China}
	\title{\Large  Scattering of charged massive scalar waves by Kerr-Newman black holes}
	
	\begin{abstract}
		The scattering of charged massive scalar waves by Kerr-Newman black holes, with incidence along the equatorial plane, is investigated in this work. The differential scattering cross section is computed using the partial wave method, with the forward divergence handled via the series reduction technique. For the first time, we systematically examine the influence of the black hole charge, electromagnetic interactions, and field mass on the equatorial cross section. Our results reveal that regardless of whether the electromagnetic interaction is present or not, the frame-dragging effect shifts the glory away from the exact backward direction and can place interference minima there, contrasting with the on-axis scattering case. The average scattered flux intensity at the medium to large scattering angles exhibits a large enhancement as the Lorentz attraction or field mass increases, particularly in the slowly rotating regime, with the enhancement being frequency-dependent. When superradiance occurs, we observe that the cross section in the prograde scattering angles ($\sim 135^{\circ} < \phi < 270^{\circ}$) increases as the black hole spin increases, due to enhanced prograde partial wave contributions. Meanwhile, the superradiant scattering cross section increases in all (except the forward) directions when the Lorentz force becomes more repulsive. These findings highlight unique equatorial-plane signatures of charged, rotating spacetimes, distinguishing them from prior on-axis analyses.  	
	\end{abstract}
	
	
	\maketitle
	\section{Introduction}
	Scattering phenomena play a pivotal role in atomic physics and elementary particle physics, as the differential scattering cross section $d\sigma/d\Omega$ in scatterings in these fields encodes detailed information about atomic potentials and particle structures \cite{Taylor:2012,Rutherford:1911,Kendall:1991}. This methodology of inferring hidden structures from scattered information has inspired analogous methods in astrophysics.  Black holes (BHs), one of the predictions of general relativity, have been strongly supported by direct imaging evidence from the Event Horizon Telescope collaborations \cite{EventHorizonTelescope:2019dse,EventHorizonTelescope:2022wkp}. However, the event horizon, as a causal boundary and null hypersurface, prevents external observers from accessing information within it using classical means. This has motivated theoretical physicists to extend scattering theory from atomic physics to the BH spacetimes, where the spacetime curvature manifests itself as an effective scattering potential. This analogy has not only opened new research avenues but also made the study of BH absorption and scattering of bosonic and fermionic fields a persistent research focus. Numerous studies focusing on the scattering of scalar field \cite{Sanchez:1977vz,Crispino:2009ki,Chen:2011jgd,Anacleto:2019tdj,Leite:2019eis,Anacleto:2020lel,Huang:2020bdf,Richarte:2021fbi,Wan:2022vcp,dePaula:2022kzz,Xavier:2023ljy,Li:2024xyu,Li:2025yoz}, Dirac field  \cite{Dolan:2006vj,Sporea:2017zxe,Sporea:2018rif,Sporea:2018rsk,Cotaescu:2018etx}, electromagnetic field \cite{Crispino:2009xt,Leite:2018mon,deOliveira:2019tlk} and gravitational field \cite{Handler:1980un,Dolan:2008kf,Crispino:2015gua,Stratton:2019deq} have been carried out. It has been found that the scattering behavior of these test fields around BHs or other compact objects gives rise to a range of fascinating physical phenomena, including but not limited to, glory \cite{Matzner:1985rjn}, orbiting scattering \cite{Anninos:1992ih}, rainbow scattering \cite{Stratton:2019deq,Leite:2019uql}, and superradiant scattering \cite{Handler:1980un,Dolan:2008kf}. However, most studies on scattering cross sections have restricted their scope to either the scattering of neutral fields in spherically symmetric spacetimes or scattering in axisymmetric spacetimes with field propagation along the rotational axis. This leaves an important, less well-understood question of the scattering, namely, how these phenomena manifest in more general scattering setups, including off-axis scattering and scattering in the presence of other kinds of interactions.
	
	In 2001, Glampedakis and Andersson \cite{Glampedakis:2001cx} investigated the off-axis scattering cross section of massless scalar waves by Kerr BHs. They found that the interference fringes are frame-dragged, preventing the glory from being detected in the exact backward direction when the massless waves propagate in the equatorial plane. To address the divergence issue arising from the long-range nature of the gravity, they decomposed the scattering amplitude into a diffraction amplitude and a Newtonian amplitude. Later, Stratton et al. \cite{Stratton:2020cps} considered the same problem but employed a method known as the \textit{series reduction method} and extended it to other bosonic fields.  They showed that both methods—the amplitude decomposition and series reduction—yield the same results for massless scalar fields. However, besides these two works, the study of scattering cross sections along off-axis directions in other rotating but non-Kerr spacetimes remains relatively underdeveloped. A deeper understanding of how various scattering phenomena behave in such scattering scenarios, especially how the BH parameters and frame-dragging affect scattering, is still needed.

	In this work, we will study the scattering cross section of a charged scalar wave along the equatorial plane by the Kerr-Newman (KN) spacetime, with particular emphasis on the effect of the BH rotation and electromagnetic interaction on the superradiance. To the best of our knowledge, previous studies on charged scalar waves around KN BHs have primarily focused on the absorption cross section rather than the scattering cross section. The absorption cross section of a massive charged scalar field around a KN BH was investigated by Benone and Crispino \cite{Benone:2019all}. Bianchi et al. \cite{Bianchi:2023lrg} derived a tree-level scattering amplitude (but not the cross section) for a charged scalar particle interacting with a KN BH. However, no attention has been paid to the scattering cross section of charged massive scalar waves incident along the equatorial plane of a KN BH in these works. We show in this work that, due to the BH spin, the glory in the equatorial scattering cross section is dragged toward the direction of spacetime rotation. Furthermore, as the Lorentz repulsion between the scalar wave and spacetime charge increases, the waves are less deflected, leading to a considerable decrease in the scattering cross section in non-forward directions. Moreover, when superradiance happens, the cross section on the side of the prograde direction will also be enhanced by the increase of the BH spin. These effects are characteristic of the equatorial scattering of charged scalar waves in rotating charged spacetimes.

	This work is organized as follows. In Sec. \ref{sec:scattering}, we briefly describe the propagation of the charged and massive scalar waves in the KN BH along the equatorial direction and give an expression for the scattering cross section by means of the partial wave method.  Sec. \ref{sec:numerical results} is devoted to presenting the numerical results of the scattering cross section and analyzing the influence of parameters. In Sec. \ref{sec:superradiance}, we discuss the effect of superradiance on the scattering cross section.  We give our concluding remarks in Sec. \ref{sec:conclusion}.  We use the natural units ($G=c=\hbar=4\pi \epsilon_{0}=1$) and the metric signature $(-,+,+,+)$  throughout this work.
	
	\section{Scattering of charged scalar field in KN BH} 
	\label{sec:scattering}
	In this section, we investigate the propagation of the charged and massive scalar waves in the KN BH to analyze the wave scattering. The KN metric in Boyer-Lindquist coordinates can be described as,
	\begin{align}\label{eq:spacetime}
		\dd s^2=&-\frac{\Delta}{\rho^2}\left(a\sin^2 \theta  \dd \phi-\dd t\right)^2 + \rho^2 \left(\frac{\dd r^2}{\Delta}+\dd \theta^2\right) \nonumber \\  &+ \frac{\sin^2\theta}{\rho^2}\left[\left(r^2+a^2\right)\dd \phi -a  \dd t\right]^2
	\end{align}
	with
	\begin{align}
		\rho^2&=r^2 +a^2 \cos^2\theta, \label{eq:rho} \\  
		\Delta&=r^2+a^2-2M r+Q^2, \label{eq:Delta}
	\end{align}
	where $M$, $a$ and $Q$ denote the mass, spin angular momentum per unit mass and charge of the BH, respectively. When $a=0$ and $Q=0$, the spacetime reduces to the Reissner-Nordstr$\ddot{\rm{o}}$m (RN)  and Kerr spacetimes, respectively. The outer event horizon is given by $\Delta=0$, namely, 
	\begin{align}\label{eq:rh}
		r_h=M+\sqrt{M^2-(a^2+Q^2)}
	\end{align}
	and  $M^2-(a^2+Q^2)>0$ needs to be  satisfied for BH parameters.
	
	In this spacetime, the evolution of the charged massive scalar field $\Psi_{\omega}$  with the incidence frequency $\omega$ is governed by  the Klein-Gordon equation, 
	\begin{align}	\label{kge}
		\left[\left(\nabla_{\alpha} - i q A_{\alpha}\right)\left(\nabla^{\alpha} - i q A^{\alpha}\right)-\mu^2\right]\Psi_{\omega} =0,
	\end{align}
	where $q$ and $\mu$ are the charge and mass of this field, respectively, and $A_{\alpha}$ is the  electromagnetic potential 
	\begin{align}\label{elec potential}
		A_{\alpha}=-\frac{Qr}{\rho^2}\left(1,0,0,-a \sin^{2}{\theta}\right).
	\end{align}
	In order to solve the above wave equation,  we apply the following  ansatz
	\begin{align}	\label{ansatz}
		\Psi_{\omega}(t,r,\theta,\phi)= \sum_{l=0}^{\infty}\sum_{m=-l}^{l}\frac{F_{\omega l m}(r)S_{\omega l m}(\theta)}{\sqrt{r^2+a^2}} \expe^ {i m \phi- i\omega t},
	\end{align}
	where $l$ and $m$ represent the angular quantum number and azimuthal quantum number, respectively.  Isolating the variables, we obtain the angular and radial wave functions, $S_{\omega l m}(\theta)$ and $F_{\omega lm}(r)$. To be more specific, the spheroidal harmonics $S_{\omega l m}(\theta)$ obey the following equation 
	\begin{align}\label{eq:spheroidal}
		&\left(\frac{\dd ^2}{\dd \theta^2}+\cot\theta\frac{\dd }{\dd \theta}\right)S_{\omega lm}\nonumber \\ 
		&+\left[\lambda_{lm}+a^2\left(\omega^2-\mu^2\right)\cos^2\theta-\frac{m^2}{\sin^2\theta}\right]S_{\omega lm}=0,
	\end{align}
	with $\lambda_{lm}$ being  the angular eigenvalue. In addition, the function $F_{\omega l m}(r) $  is subject  to the following second-order differential equation
	\begin{align}\label{eq:radial}
		\left(\frac{\dd ^2}{\dd  r_{*}^2}+V_{\omega lm}\right)F_{\omega lm}(r_{*})=0,
	\end{align}
	with $r_{*}$ being   the tortoise coordinate,
	\begin{align}\label{eq:tortoisel}
		r_{*}\equiv\int \dd r \left(\frac{r^2+a^2}{\Delta}\right),
	\end{align}
	and $V_{\omega lm}$ being  the  potential \cite{Benone:2019all},
	\begin{widetext}
		\begin{align}\label{eq:veff}
			V_{\omega lm}(r)&=\frac{\left[\left(r^2+a^2\right)\omega - am - q Qr\right]^2+ \left[2ma\omega-\mu^2 \left(r^2+a^2\right) - \lambda_{l m}-a^2\left(\omega^2-\mu^2\right)\right]\Delta}{\left(r^2+a^2\right)^2}\nonumber \\  
			&-\frac{\left[\Delta+2r\left(r-M\right)\right]\Delta}{\left(r^2+a^2\right)^3} + \frac{3r^2\Delta^2}{\left(r^2+a^2\right)^4}.
		\end{align}
	\end{widetext}
	By analyzing this potential at the horizon  and infinity, we have  
	\begin{align}  
		\omega_{h} &\equiv \sqrt{V_{\omega lm}(r_h)}=\omega - \frac{a m+q Q r_h}{r_h^2+a^2}\equiv  \omega- \omega_c, \label{eq:omega_c}  \\  
		\omega_\infty& \equiv \sqrt{V_{\omega lm}(\infty)}=\sqrt{\omega^2-\mu^2}, \label{eq:omega_inf}
	\end{align}
   where, in the second equality of Eq.~\eqref{eq:omega_c}, we define $\omega_c$ as the critical frequency.
	
	Considering that we are concentrating on the scattering problem of the charged massive scalar field, i.e., the scattering wave is unbound, the condition $\omega>\mu$ has to be fulfilled. The asymptotic solutions of Eq. \eqref{eq:radial} are of the following forms \cite{Futterman:1988}
	\begin{align}\label{eq:solution}
		F_{\omega lm}(r) \approx
		\left\{
		\begin{array}{ll}
			\expe^ {-i\omega_\infty r_{*}} + \mathcal{R}_{\omega l m} \expe^ {i\omega_\infty r_{*}}, \quad &\mbox{for $r_{*}\rightarrow +\infty$},\\
			\mathcal{T}_{\omega l m}   \expe^ {-i \omega_{h} r_{*}}, &\mbox{for $r_{*}\rightarrow -\infty~$},
		\end{array}
		\right.
	\end{align}
	where $\mathcal{R}_{\omega l m}$ and $\mathcal{T}_{\omega l m}$ represent the reflection and transmission coefficients, respectively, and are determined by the following relationship
	\begin{align}\label{eq:rtf}
		|\mathcal{R}_{\omega lm}|^2 + \frac{\omega_{h}}{\omega_\infty}|\mathcal{T}_{\omega lm}|^2 = 1.
	\end{align}
	Obviously, when  $\omega_{h} <0$, superradiance occurs, i.e.,    $|\mathcal{R}_{\omega lm}|^2> 1$.
	
	Using the partial wave method, the differential scattering cross section  (hereafter referred to as ``cross section" for simplicity) along the  outgoing direction $(\theta,\,\phi)$ from a charged massive scalar field impacting the BH from the incoming direction $(\gamma,\phi_0)$ is given by
	\cite{Futterman:1988}
	\begin{align} \label{eq:dscs}
		\frac{\dd \sigma}{\dd \Omega} =\left|f(\theta,\phi)\right|^2,
	\end{align}
	with the scattering amplitude $f(\theta,\phi)$ taking the form
	\begin{align}\label{eq:sa}
		f(\theta,\phi)& =\frac{2\pi}{i\omega_\infty}\sum_{l=0}^{\infty}\sum_{m=-l}^{l}S_{\omega lm}(\gamma)S_{\omega lm}(\theta) \nonumber \\
		& \times\expe^{im(\phi-\phi_0)}\left[(-1)^{l+1}\mathcal{R}_{\omega lm}-1\right],
	\end{align}
	where the $S_{\omega l m}$ are calculated using Eq. \eqref{eq:spheroidal}. In the next section, we will examine the numerical results for the cross section \eqref{eq:dscs} and analyze how it is related to the BH and field parameters.
	
	\section{numerical results and analysis}\label{sec:numerical results}
	
   In this section, we present and discuss the results of the cross section (\ref{eq:dscs}) calculated using the partial wave method. To obtain the scattering amplitude \eqref{eq:sa}, we have to calculate the reflection coefficient $\mathcal{R}_{\omega lm}$ and the spin-0 spheroidal harmonics $S_{\omega l m}$. The former is obtained by solving the radial wave equation \eqref{eq:radial} using the fourth-order Runge-Kutta method for optimal accuracy, with the initial values detailed in Ref. \cite{Dolan:2009zza}. The spin-0 spheroidal harmonics, which satisfy the angular equation \eqref{eq:spheroidal} and whose corresponding eigenvalues are $\lambda_{lm}$, are computed by the spectral decomposition method \cite{Hughes:1999bq} with the help of the BH Perturbation Toolkit \cite{Toolkit}. In addition, we adopt a series reduction technique to efficiently handle the infinite summation in the scattering amplitude (\ref{eq:sa}) \cite{Stratton:2020cps}. Finally, we set $M = 1$ when plotting the figures, which implies that the numerical values of the BH parameters $(a,\,Q)$ are scaled by $M$, while the field parameters $(\mu,\,q,\,\omega)$ are scaled by $M^{-1}$.

	\begin{figure}[htp!]
		\centering
		\includegraphics[width=0.5\textwidth]{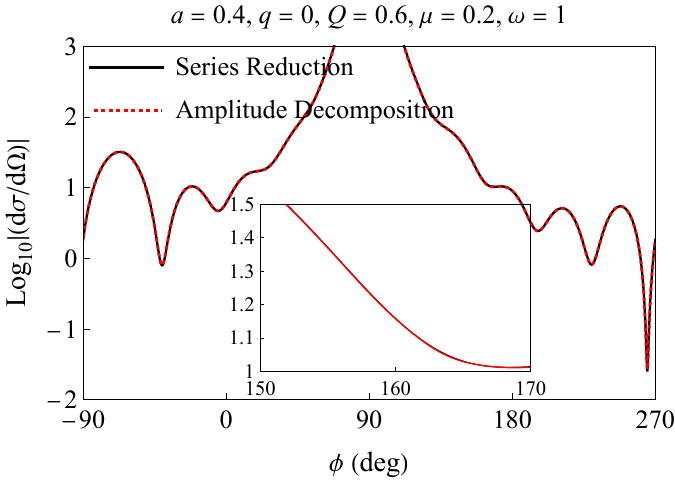} 
		\includegraphics[width=0.5\textwidth]{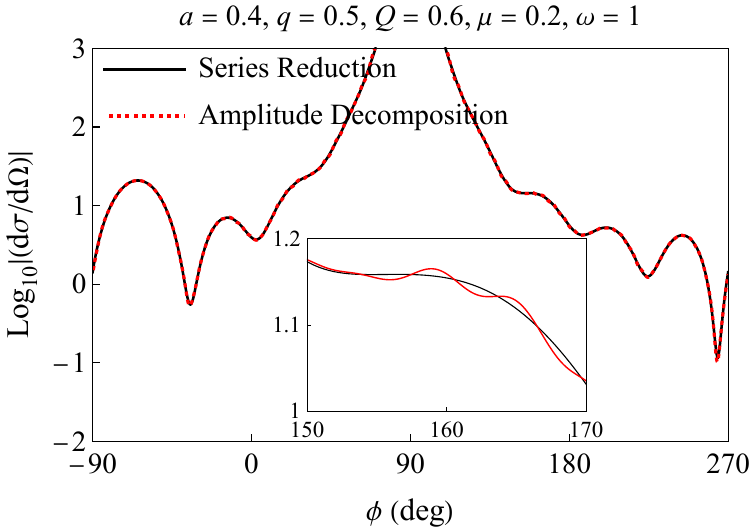}
		\caption{Comparison of the cross sections at $\theta=\pi/2$, computed using the series reduction and amplitude decomposition methods, for the incoming direction $(\gamma, \phi_0) = (\pi/2, \pi/2)$. The insets show the cross sections in the range $150^\circ < \phi < 170^\circ$  to highlight the differences more clearly.}
		\label{fig:com}
	\end{figure}

	To ensure the reliability of our numerical results, we first compare in Fig.~\ref{fig:com} the cross sections obtained via two methods for computing the scattering amplitudes, the series reduction method \cite{Stratton:2020cps} and the amplitude decomposition approach \cite{Glampedakis:2001cx}. It is seen from the top plot of Fig.~\ref{fig:com} that when $q=0$, i.e., the electromagnetic interaction is turned off, the cross section from these two methods agrees perfectly. When $q>0$ is present, the inset of the bottom plot shows that the cross section obtained by the amplitude decomposition method exhibits slight oscillations in certain $\phi$ ranges, although the two methods are generally still consistent. Therefore, in the following studies, we will use only the series reduction method to compute the cross sections.

	\begin{figure}[htp!]
		\centering
		\includegraphics[width=0.45\textwidth]{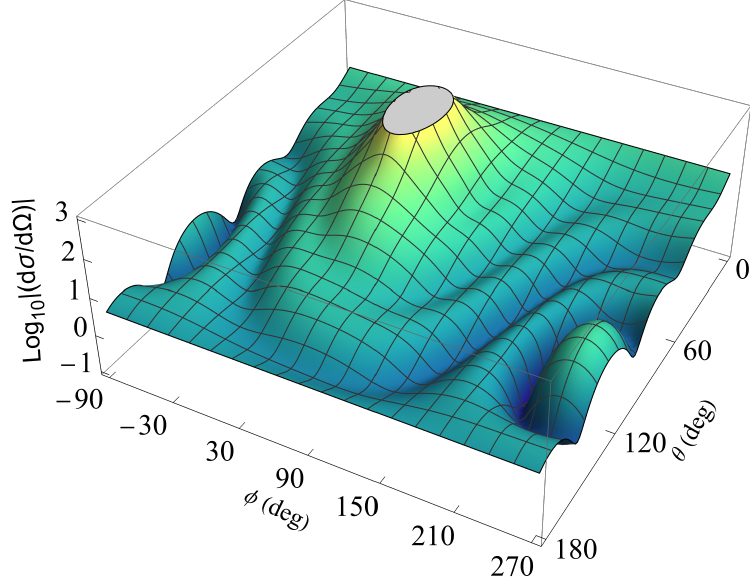}
		\includegraphics[width=0.45\textwidth]{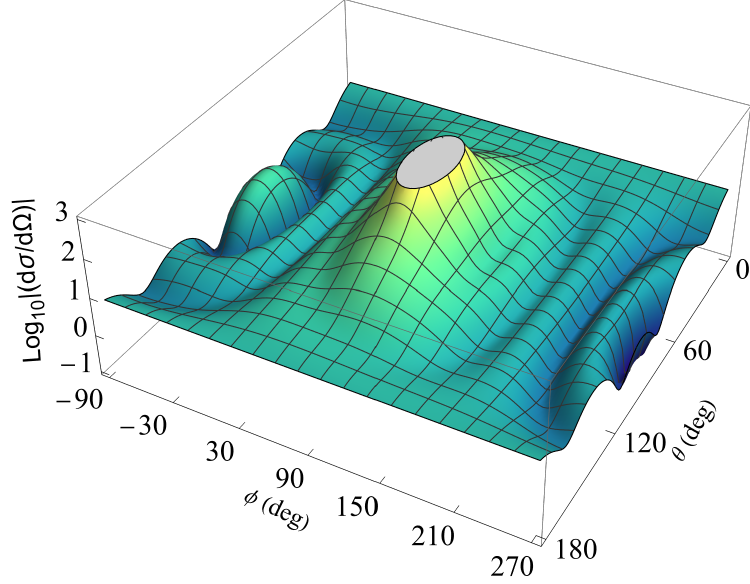}
		\caption{The cross sections for fixed parameters $a=0.4,\, q=0.5,\, Q=0.6,\,\mu=0.2$ and $\omega=1$, for scalar waves incident from $(\gamma,\phi_0)=(\pi/3,\pi/4)$ (top) and $(\pi/2,\pi/2)$ (bottom).}
		\label{sca_3D}
	\end{figure}
	
	Before a detailed analysis of the cross section and its dependence on BH and field parameters, we first present in Fig.~\ref{sca_3D} a general study of the cross sections for some exemplary parameter values, considering general $\theta$ and $\phi$. This figure shows two cross sections for incident waves propagating along $(\gamma, \phi_0) = (\pi/3, \pi/4)$ (top) and $(\pi/2, \pi/2)$ (bottom), with fixed parameters $a=0.4,\, q=0.5,\, Q=0.6,\,\mu=0.2$, and $\omega=1$. First, we observe divergences in both cross sections along the incident direction due to the long-range nature of both the gravitational and electromagnetic interactions. Second, for waves propagating in the non-equatorial plane (top), the corresponding cross section exhibits no symmetry with respect to $\phi$. This asymmetry is expected due to the presence of a preferred direction along the spacetime spin and the fact that the incident wave is neither along nor perpendicular to it. In contrast, the cross section for a wave incoming along the equatorial direction (bottom) possesses reflection symmetry about the equatorial plane ($\theta=\pi/2$).

	In the following subsections, to better understand the influence of frame-dragging induced by $a$ and other parameters on the cross section, our investigation will be categorized into three scenarios: (\ref{subsec:uncharged}) the neutral scalar wave scattered by a KN BH; (\ref{subsec:sr}) the charged scalar wave scattered by a slowly rotating KN BH with $0 \leq |a|\leq 0.4$; and (\ref{subsec:rr}) the charged scalar wave scattered by a rapidly rotating KN BH with $0.5\leq |a|\leq 1$.

	\subsection{Scattering of neutral scalar wave by KN BH}
	\label{subsec:uncharged}
	
	\begin{figure*}[htp!]
		\centering
		\includegraphics[width=0.5\textwidth]{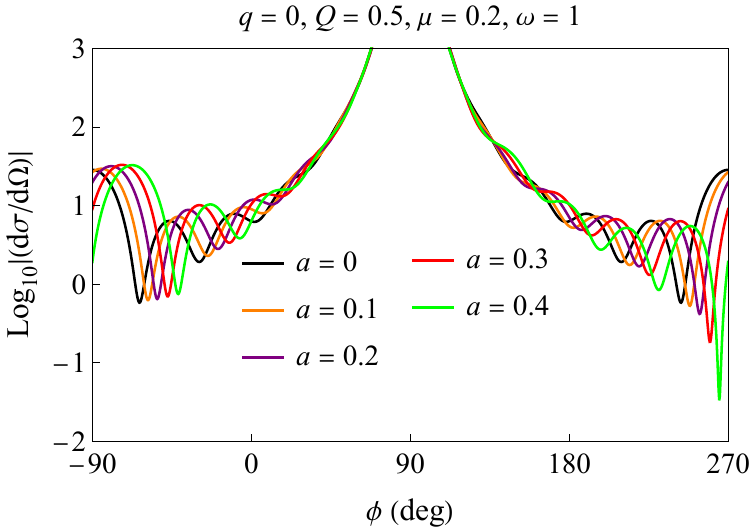}\includegraphics[width=0.5\textwidth]{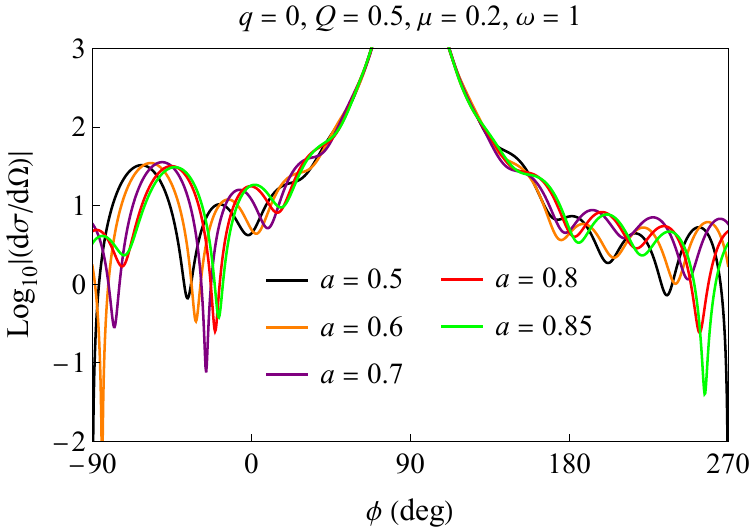}
		
		\caption{The cross section as a function of $\phi$ at fixed $\theta = \pi/2$ for neutral massive scalar fields scattered by a KN BH with different values of $a$. The incident wave propagates along $(\gamma, \phi_0) = (\pi/2, \pi/2)$. Left: slowly rotating case; right: rapidly rotating case. }
		\label{fig:sca_a}
	\end{figure*}
	
   For this case, we present in Fig.~\ref{fig:sca_a} the cross section as a function of $\phi$ at the specific angle $\theta=\pi/2$ for various values of $a$, for a neutral massive scalar wave propagating along the direction $(\gamma,\phi_0)=(\pi/2,\pi/2)$. This orientation ensures that the partial waves experience the strongest rotational effects, and all subsequent discussions are based on this setup.

   For the effect of $a$, we first notice that, due to the freedom in the definitions of the rotation direction and $\phi$-direction, the cross section for $a<0$ is related to that for $a>0$ by a straightforward relation
  \begin{equation}\label{eq:antisym}
	\frac{\dd \sigma}{\dd \Omega}^{\text{prograde}}(\phi,a>0) =\frac{\dd \sigma}{\dd \Omega}^{\text{retrograde}} (\pi-\phi,a<0).
  \end{equation}
  This symmetry effectively reduces the independent parameter space, and therefore we only need to concentrate on the case $a\geq 0$ with the full range of $\phi\in[-\pi/2,3\pi/2]$ without loss of generality.

	Firstly, we consider the slowly rotating case ($|a|<0.5$) \cite{Myung:2021fzo}, whose rotational effect is relatively weak. In the left panel of Fig.~\ref{fig:sca_a}, we see that the cross sections are increasingly shifted toward the spacetime spin direction as the BH spin increases, resulting in an asymmetry with respect to the incident direction $\phi_0=\pi/2$. This effect is similar to the scattering in pure Kerr spacetime reported in Ref. \cite{Glampedakis:2001cx}. As the parameter $a$ increases, the amplitudes of a series of interference maxima, including the glory peak (the largest of these maxima), are also frame-dragged along the BH rotation direction to larger $\phi$. Moreover, we observe that the oscillations of the cross section intensify as $\phi$ moves from the incident direction toward the glory peak.

	In the rapidly rotating cases ($0.5 \leq |a| \leq 1$, Fig.~\ref{fig:sca_a} right panel), it is seen that compared to the $|a|<0.5$ case, both the interference fringes and the glory peak locations are further shifted toward larger $\phi$. Due to this continuous shift as $a$ increases, the global minima of the cross sections may now occur exactly backward $(\phi=-\pi/2)$, as seen from the $a=0.5$ case (black curve). This is in contrast to the case of a static BH ($a=0$) or a rotating BH with the test wave propagating along the rotation axis \cite{Li:2025yoz}.  Moreover, when $|a|\geq 0.5$, we observe that the oscillation amplitudes of the cross section before $\phi$ reaches the glory peak location from below reduce as $a$ increases. This differs from the behavior observed in the slowly rotating case. Furthermore, due to the strong frame-dragging, the cross section peaks and minima between $\phi=90^\circ$ and $\phi=270^\circ$ show non-monotonic behavior as $a$ increases. Finally, it is worth mentioning that because of the frame-dragging effect, we cannot directly observe the influence of the rotation parameter $a$ on the width of the interference fringes, as in the case of incidence along the axis \cite{Leite:2019eis}.

	\begin{figure*}[htp!]
		\centering
		\includegraphics[width=0.5\textwidth]{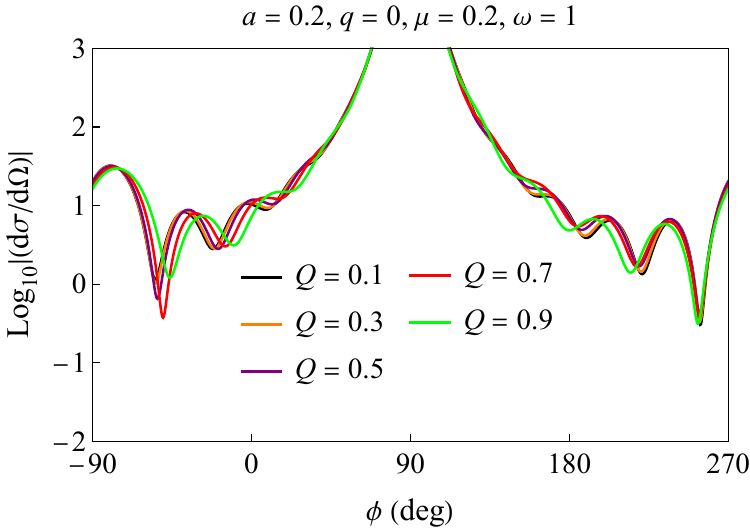}\includegraphics[width=0.5\textwidth]{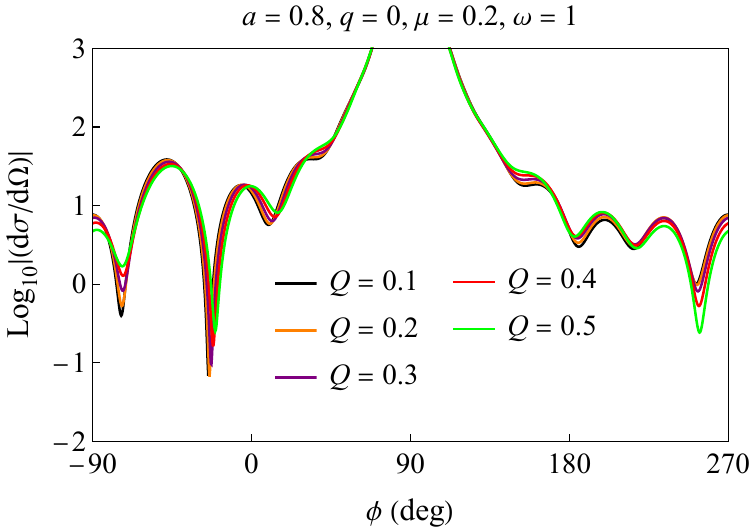}
		\caption{The cross section as a function of $\phi$ at fixed $\theta=\pi/2$ for neutral massive scalar fields scattered by a KN BH with different values of $Q$. The incident wave propagates along $(\gamma, \phi_0) = (\pi/2, \pi/2)$. Left: slowly rotating case; right: rapidly rotating case.}
		\label{fig:sca_Q}
	\end{figure*}
	
	When a neutral scalar wave propagates along the rotation axis and hits the KN BH, Leite et al. \cite{Leite:2019eis} showed that as the BH charge $Q$ increases, the interference fringes become wider. For a neutral wave propagating on the equatorial plane in a slowly rotating KN BH, we observe in the left panel of Fig.~\ref{fig:sca_Q} that the width of the interference fringes also increases as the BH charge $Q$ increases. This is understandable because the parameter $Q$ does not break spherical symmetry for the neutral wave, and therefore its effect on the cross section should be independent of the incident direction. These effects are consistent with the influence of charge $Q$ on wave scattering in other charged spacetimes \cite{Li:2024xyu,Li:2025yoz}.  	For the rapidly rotating case shown in the right panel, a closer inspection reveals that the broadening of the interference fringes with larger $Q$ also occurs, although it is less apparent than in the slower rotation case.

	From both plots of Fig.~\ref{fig:sca_Q}, we note that the parameter $Q$ enhances the broadening of the interference fringes more significantly for directions further from the glory peak angle around $-90^\circ$ (or equivalently $270^\circ$). For peaks roughly at $\phi\approx 200^\circ$, this broadening shifts them toward smaller $\phi$ for larger $Q$, in contrast to the effect of $a$, which drags the peaks rightwards. 
	Regarding the effect of $Q$ on the absolute value of the cross section, we observe that, roughly speaking, larger $Q$ tends to increase the total cross section in the forward direction ($0^\circ\lesssim \phi\lesssim 180^\circ$) and reduce it in the backward direction ($180^\circ \lesssim \phi\lesssim 360^\circ$). This is consistent with the effect of charge $Q$ on the BH shadow size in RN and KN spacetimes \cite{Crispino:2009ki, Pang:2018jpm,Babar:2020txt}. That is, the larger the $Q$, the smaller the shadow and therefore the larger the forward total cross section.

	\subsection{Charged scalar wave scattering by slowly rotating BH}
	\label{subsec:sr}
	
	\begin{figure*}[htp]
		\centering
		\includegraphics[width=0.5\textwidth]{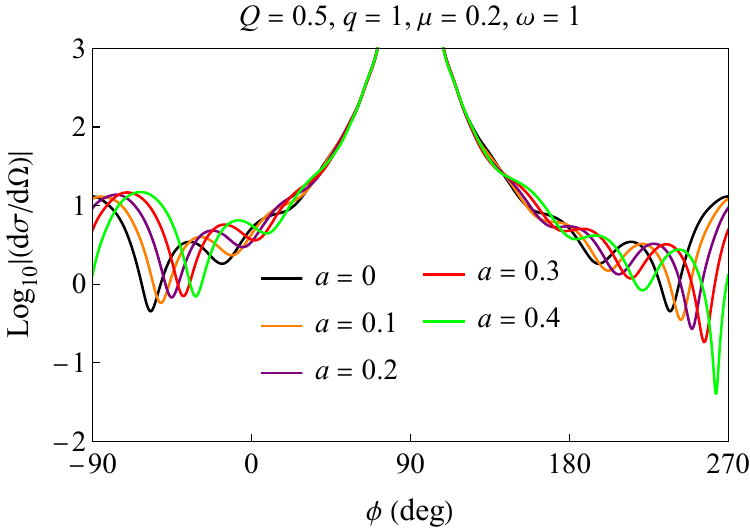}\includegraphics[width=0.5\textwidth]{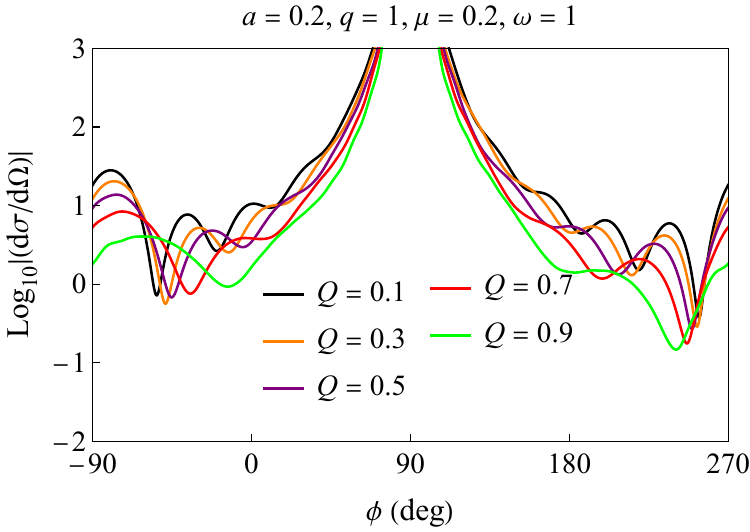}
		\includegraphics[width=0.5\textwidth]{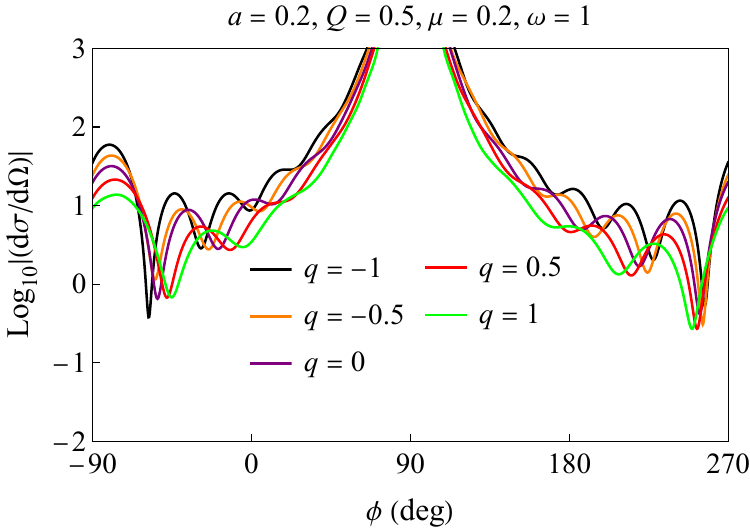}\includegraphics[width=0.5\textwidth]
		{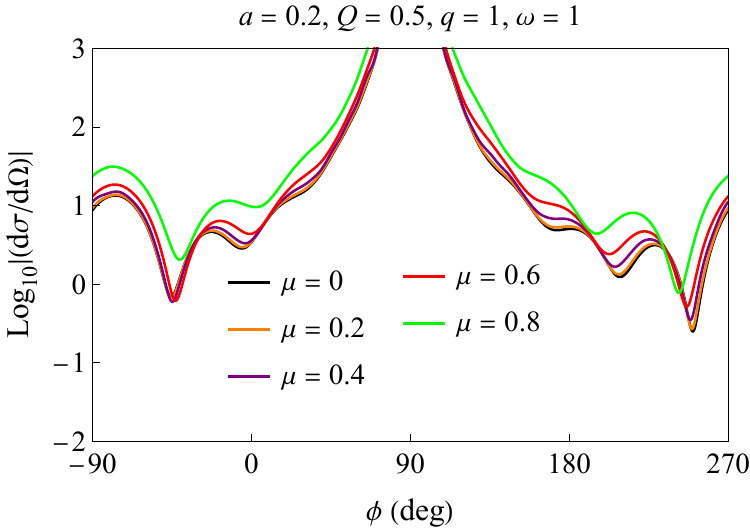}
		\includegraphics[width=0.5\textwidth]{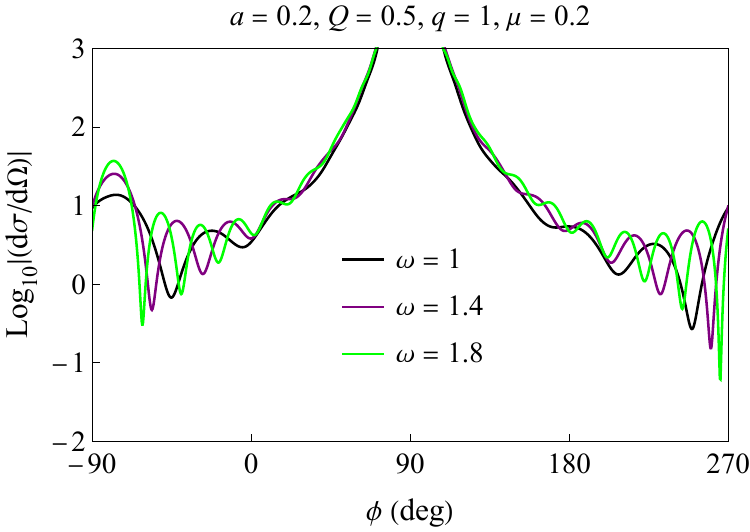}\includegraphics[width=0.5\textwidth]{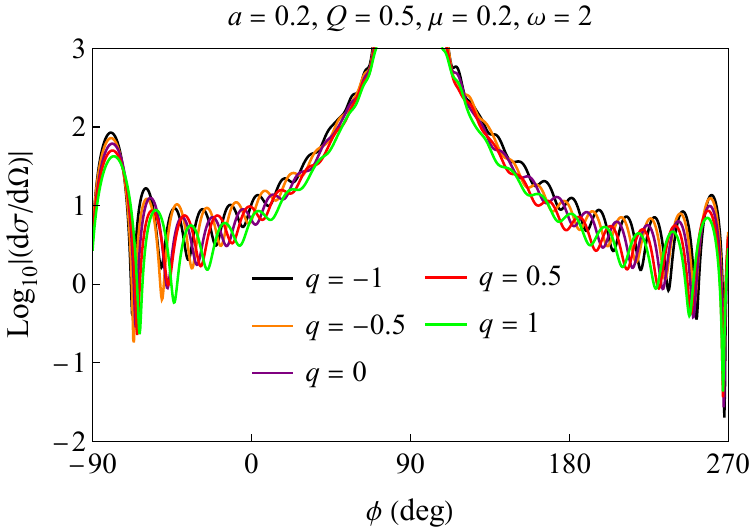}
		\caption{The cross section as a function of $\phi$ at fixed $\theta=\pi/2$ for a charged massive scalar field scattered by a slowly rotating KN BH, shown for different values of $a$ (top left), $Q$ (top right and bottom right), $q$ (middle left), $\mu$ (middle right), and $\omega$ (bottom left). The incident wave propagates along $(\gamma, \phi_0) = (\pi/2, \pi/2)$.}
		\label{fig:sca_q}
	\end{figure*}

Now, we consider the case of a charged massive scalar wave scattered by a slowly rotating BH and analyze the effects of the parameters $(a,\,Q,\,q,\,\mu,\,\omega)$ as shown in Fig.~\ref{fig:sca_q}. For the parameters $Q$ and $q$, since the gravitational dynamics only depend on the size of $Q$,  while the electromagnetic interaction depends on the sign of $qQ$, we may focus on the case $Q \geq 0$ with both signs of $q$ without loss of generality.

From Fig.~\ref{fig:sca_q}, we see that when the electromagnetic interaction is present, cross sections from all plots exhibit the frame-dragging effect towards the rotation direction, which is similar to the neutral case in Fig.~\ref{fig:sca_a}. However, in this case, the frame-dragging effect is coupled with the electromagnetic interaction, as can be seen from the radial equation of particle motion \cite{Carter:1968rr}. 
By carefully comparing the left panel of Fig. \ref{fig:sca_a} with the top left panel of Fig. \ref{fig:sca_q}, we can see that the dependence of the interference extrema of the cross sections on the spin parameter $a$ exhibits a slight change, which can be attributed to the electromagnetic interaction because other parameters except $q$ in these two scatterings are the same.

From the top right and middle left panels of Fig.~\ref{fig:sca_q}, we observe that increasing $Q$ and/or $q$, thereby strengthening the repulsive electromagnetic interaction between the BH and the field, leads to a noticeable broadening of the interference fringes in the backward direction. At the same time, the average scattering flux intensities at medium-to-large $\phi$ angles are significantly suppressed. This behavior aligns with the intuitive expectation that a stronger repulsive interaction, competing against gravitational attraction, results in weaker deflection and thus a reduced flux toward larger scattering angles while enhancing forward scattering. Moreover, the oscillation amplitude of the cross section (i.e., the contrast between maxima and minima) decreases as $Q$ or $q$ increases. These qualitative features are consistent with our previous findings for charged Horndeski BHs~\cite{Li:2024xyu} and with scattering by KN black-bounce spacetime along the rotation axis~\cite{Li:2025yoz}. However, they were less pronounced there due to the higher wave frequency ($\omega=2$) used. In contrast, the frequency used here is comparable to the field mass, making the cross section more sensitive to variations in the electric potential. This behavior is also consistent with general expectations from quantum electrostatic scattering, such as in Mott scattering~\cite{Peskin:2018}.

Regarding the influence of the field mass, the middle right panel clearly demonstrates that increasing the field mass leads to broader interference fringes and enhanced average scattering flux intensities. From a physical perspective, increasing the field mass while fixing $\omega$ effectively reduces the kinetic energy per unit mass of the wave, causing it to move more sluggishly in the potential. As a result, the wave is more easily deflected by gravity and leads to an enhancement of the cross section. This trend is also shown in our previous analysis of the massive scalar field scattered by the charged Horndeski BH \cite{Li:2024xyu}, where both the deflection angle and the classical differential cross section increase with the field mass. From the mathematical point of view, this can be understood from the scattering amplitude in Eq.~\eqref{eq:sa}, because as the field mass approaches the wave frequency (i.e., $\omega_\infty \to 0$), the $1/\omega_\infty$ term grows rapidly, thereby amplifying the scattering intensity. For the effects of the wave frequency $\omega$, we notice from the bottom two plots that as $\omega$ increases, there are more fringes in the oscillation pattern while the width of the fringes narrows. This behavior arises because the phase difference between different scattering paths varies more rapidly at higher frequencies, leading to finer interference patterns and stronger constructive interference, which enhances the oscillation amplitude.

The last plot of Fig.~\ref{fig:sca_q} illustrates the influence of $q$ on the cross section with respect to the wave frequency $\omega$. Compared with the middle left panel, it is evident that the effect of $q$ becomes significantly weaker at larger values of $\omega$. This indicates that $\omega$ serves as an energy scale that modulates the contribution of other energetic parameters, such as $q$, which controls the electric interaction strength. Indeed, we also generated a similar plot analyzing the influence of the field mass $\mu$ at a higher $\omega$ than that used in the middle right panel, and a similar reduction in the effect of $\mu$ was observed. This dependence of the influence of the electric force and field mass on the cross section with respect to $\omega$ is also present in the \textit{absorption} cross section. As reported in Ref.~\cite{Benone:2015bst} (see Figs.~3 and 4 therein), the impact of the electric force and field mass on the absorption cross section gradually diminishes as the frequency increases. This behavior is physically reasonable: at low frequencies, the absorption process is sensitive to the potential barriers (\ref{eq:veff}) near the BH, which include both gravitational and electromagnetic contributions; whereas at high frequencies, the wave is less affected by these barriers and falls directly into the BH, such that the absorption cross section becomes dominated by the geometric properties of the spacetime. Therefore, our results suggest that detecting waves with longer wavelengths is preferable when attempting to probe the effects of parameters such as $(q,\,Q)$ and $\mu$ on either the scattering or absorption cross sections.

\subsection{Charged scalar wave scattering by rapidly rotating BH}
\label{subsec:rr}

\begin{figure*}[htp]
	\centering
	\includegraphics[width=0.5\textwidth]{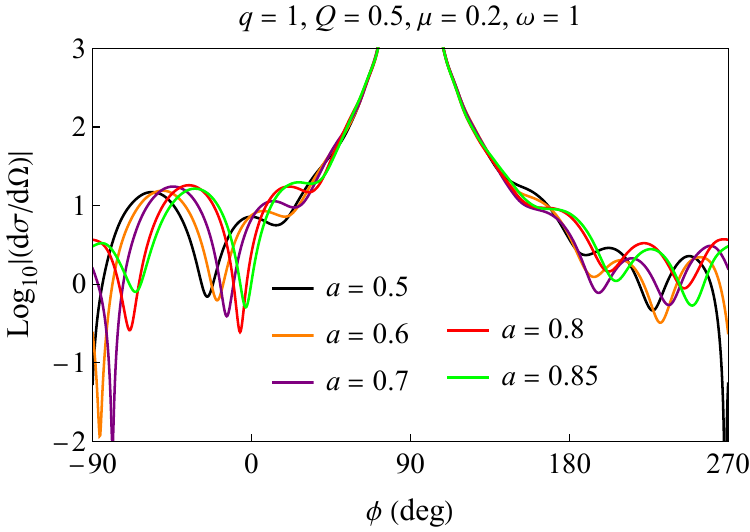}\includegraphics[width=0.5\textwidth]{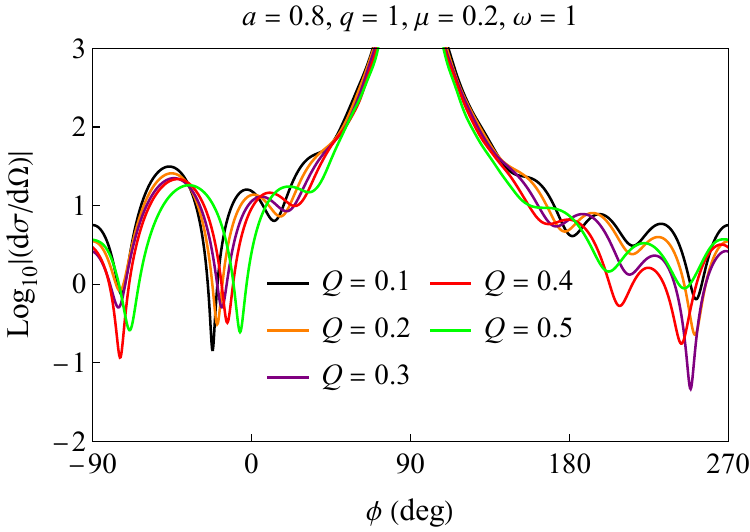}
	
	\includegraphics[width=0.5\textwidth]{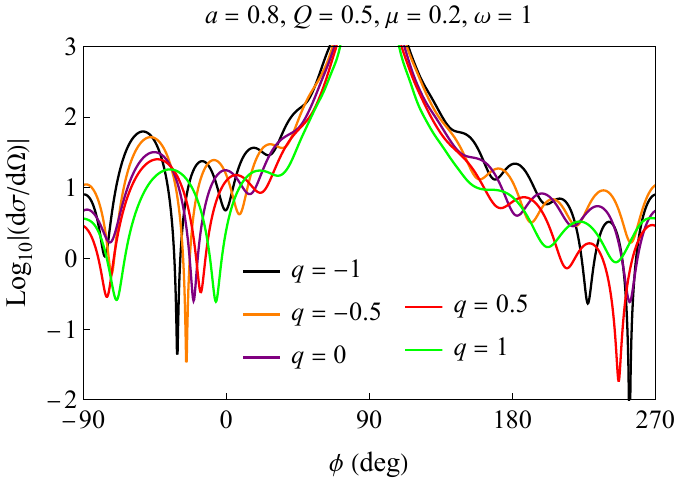}\includegraphics[width=0.5\textwidth]
	{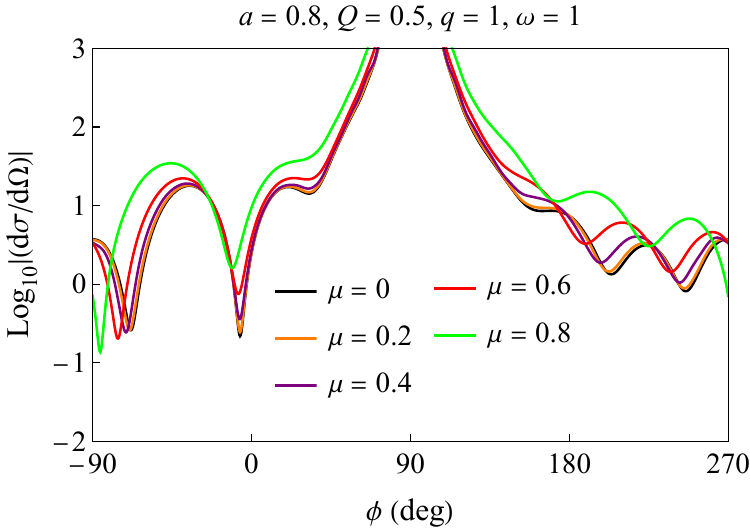}
	
	\includegraphics[width=0.5\textwidth]{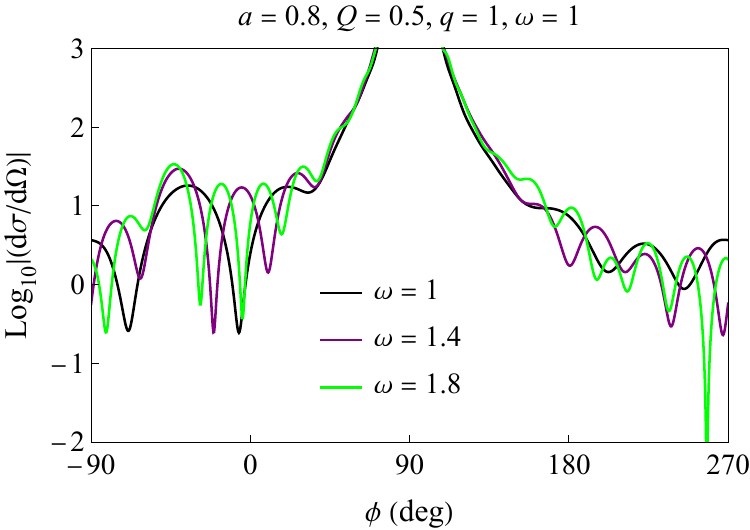}\includegraphics[width=0.5\textwidth]{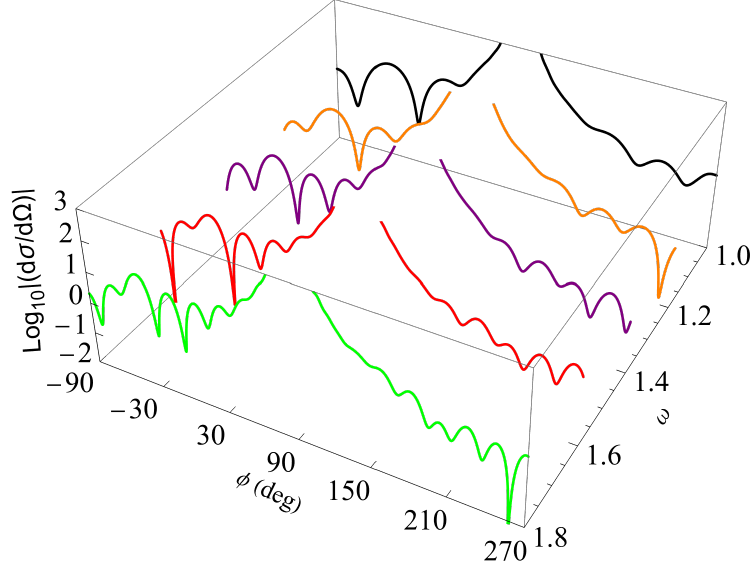}
	\caption{The cross section as a function of $\phi$ at fixed $\theta=\pi/2$ for a charged massive scalar field scattered by a  rapidly rotating KN BH, shown for different values of $a$ (top left), $Q$ (top right), $q$ (middle left), $\mu$ (middle right) and $\omega$ (bottom). The incident wave propagates along $(\gamma, \phi_0) = (\pi/2, \pi/2)$.}
	\label{fig:sca_aq}
\end{figure*}

Similar to the previous subsection, in Fig.~\ref{fig:sca_aq} we investigate the effect of the parameters $(a,\, Q,\, q,\, \mu,\, \omega)$ on the cross section for a rapidly rotating BH. In the top left panel, we observe that the cross section is shifted as the BH spin increases due to frame-dragging. Compared to the case of a rapidly rotating BH without electromagnetic interactions (right panel of Fig.~\ref{fig:sca_a}), we can clearly see that the cross section exhibits a minimum that does not occur in the backward direction for the case of $a=0.5$, as well as a smaller amplitude of the oscillations due to electromagnetic interactions. Otherwise, there is no qualitative change.  
When the other parameters are fixed and the BH charge increases, as shown in the top right panel, the differences between the cross sections become larger in the angular range $\sim180^\circ < \phi < \sim 240^\circ$, compared with the rapidly rotating case without electromagnetic interactions (right panel of Fig.~\ref{fig:sca_Q}). This behavior occurs because increasing $Q$ enhances  repulsive electromagnetic interactions. Moreover, the overall trend of the cross sections has changed. Since stronger frame-dragging leads to further deformation in the cross section, we observe in the top right and middle left panels that there is no reduction in the scattered flux intensity, particularly in the range $90^\circ < \phi < 270^\circ$, when the BH charge or the field charge is increased, i.e., when repulsive electromagnetic interactions become stronger. This result contrasts with the slowly rotating case under repulsive electromagnetic effects (middle left panel of Fig.~\ref{fig:sca_q}).
Furthermore, when other parameters are held fixed and the field mass is increased, we still observe a corresponding enhancement in the intensity of the scattered flux, except for a slight deformation of the cross section, as shown in the middle right panel.

To better understand the influence of the incident frequency, we present the dependence of the cross section on the frequency using both 2D and 3D visualizations in the bottom panels. The 2D plot allows for a direct comparison of the cross section amplitudes, while the 3D plots illustrate the continuous variation of the cross section as the incident frequency changes. We observe that both the oscillatory behavior of the cross section and the width of the interference fringes change with frequency. As the frequency increases, the oscillations become more pronounced and the fringes become narrower. In addition, the deformation of the cross section becomes increasingly significant, particularly within the range $90^\circ < \phi < 270^\circ$. 

\section{Superradiance}\label{sec:superradiance} 
In this section, we discuss the variation of the cross section  (\ref{eq:dscs}) when superradiance happens.  The superradiance condition for charged scalar fields in a charged rotating  BH background is given by $\omega_h<0$ in Eq. \eqref{eq:omega_c}, meaning that
\begin{equation}\label{eq:sur}
	\omega < \frac{a m + q Q r_h}{r_h^2 + a^2} =\omega_c.
\end{equation}
This inequality implies that both Kerr and RN BHs can exhibit superradiance when scattering charged scalar waves. Moreover, it indicates that spacetimes with nonzero spin ($a \neq 0$) can also produce superradiance even for neutral waves ($q=0$). Regardless of whether $q$ vanishes or not, we find that for prograde partial waves with $\mathrm{sign}(am)>0$ (or retrograde waves with $\mathrm{sign}(am)<0$) and a fixed frequency $\omega$, increasing (or decreasing) the spin parameter makes the superradiant condition easier to satisfy, thereby leading to stronger superradiance. However, the degree of superradiance for neutral and charged fields in spinning spacetimes can differ significantly. For the Kerr BH, the maximal amplification of neutral scalar waves is only about $0.4\%$, meaning that superradiance has no substantial observable effect on the cross section. In contrast, as demonstrated by Brito et al.~\cite{Brito:2015oca}, when the field charge $q$ is sufficiently large, the amplification factor can approach $100\%$. In the latter case, the superradiance clearly will induce measurable modifications to the cross section. Our previous works \cite{Li:2024xyu,Li:2025yoz} on charged scalar wave scattering revealed that superradiance correlates with enhanced scattered flux intensity at specific scattering angles. To further investigate these effects on the cross section, we study a physical scenario of a slowly rotating BH background with stronger repulsive electromagnetic interactions, thereby enhancing the superradiant effects to substantially higher levels.

\begin{figure*}[htp]
	\centering
	\includegraphics[width=0.5\textwidth]{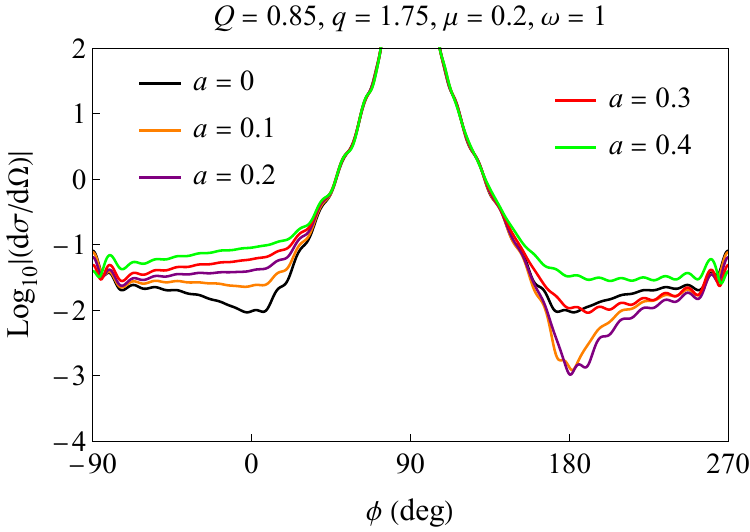}\includegraphics[width=0.5\textwidth]{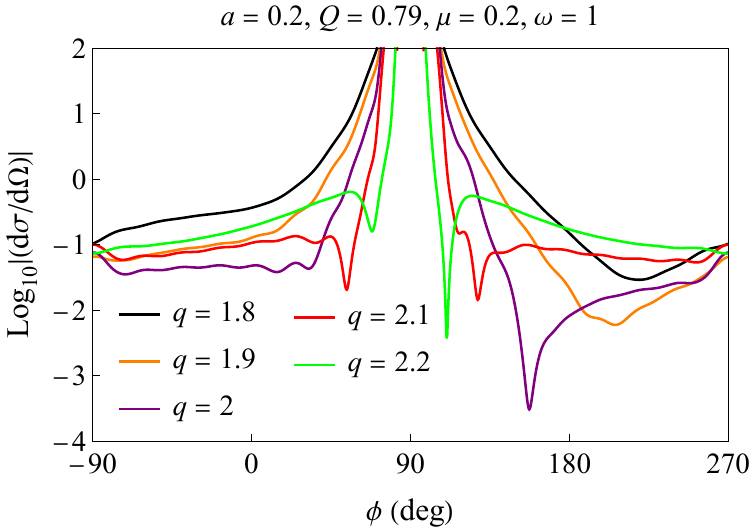}
	\caption{The cross section as a function of $\phi$ at fixed $\theta = \pi/2$ of a slowly rotating KN BH for different values of $a$ (left) and $q$ (right). The incident  wave propagates  along ($\gamma, \phi_0) = (\pi/2, \pi/2)$.}
	\label{fig:sur_sca}
\end{figure*}

In Fig.~\ref{fig:sur_sca}, we present the cross section as a function of $\phi$ at $\theta = \pi/2$ for different values of $a$ (left) and generally larger $q$ (right) compared with previous figures, where the incident waves propagate in the ($\gamma, \phi_0) = (\pi/2, \pi/2)$ direction.   
In Fig.~\ref{fig:sca_q} (top left), we can tell that the average intensity of the scattered flux tends to decrease in the range $\sim 135^\circ < \phi < 270^\circ$ as the rotation parameter increases, while it appears to increase in the range $-90^\circ < \phi <\sim 45^\circ$. However, in Fig.~\ref{fig:sur_sca} (left), where we adopt larger values of $q$ and $Q$ than those used in Fig.~\ref{fig:sca_q} (top left), such that the electric repulsion becomes significantly stronger, we can see that the dependence of the cross section on $a$ exhibits more complex behavior. The cross section in the range $\sim 135^\circ < \phi < \sim 270^\circ$ first decreases as $a$ increases from zero to about $0.1\sim0.2$ and then increases as $a$ further increases to $0.4$. Whereas in  the $\sim -90^\circ < \phi < \sim 45^\circ$ range, the cross section keeps increasing as $a$ increases. 

These features actually can be well explained by the superradiance. For the cross section in the range $90^\circ < \phi < 270^\circ$ (or the range $-90^\circ < \phi < 90^\circ$), it is known that the prograde (or retrograde) partial waves with $\mathrm{sign}(am)>0$ (or $\mathrm{sign}(am)<0$) are scattered into this region \cite{Syu:2025rex}, if frame-dragging is completely ignored. Therefore, one would expect, from Eq. \eqref{eq:sur}, that increasing $a > 0$ makes it easier for the $m>0$ partial waves (those contributing to $90^\circ < \phi < 270^\circ$) to satisfy the superradiant condition and therefore stronger superradiance, i.e, the larger cross section, happens.  Conversely, by a similar argument, the superradiance/cross section in the $-90^\circ < \phi < 90^\circ$ range, contributed more by the $m<0$ partial waves, should decrease as $a(>0)$ increases. However, due to frame-dragging, a positive $a$ tends to shift part of the $m>0$ partial waves into the $-90^\circ < \phi < 90^\circ$ region. As $a$ increases, more such waves are dragged into this range, causing the cross section there to increase rather than decrease. Clearly, at this stage, the variation of superradiance should have been dominantly controlled by the frame-dragging effect rather than the natural splitting of $m>0$ and $m<0$ partial waves into the two angular ranges. 

From the middle left panel of Fig.~\ref{fig:sca_q}, we observe that the average intensity of the scattered flux at medium to large $\phi$ angles decreases as the repulsive electromagnetic interaction increases. Therefore, we expect that the scattered flux intensity will continue decreasing until the onset of superradiance determined by condition~\eqref{eq:sur}, and will subsequently start increasing due to the superradiance effect as the field charge, i.e., the repulsive electromagnetic interaction, is further increased. For the given parameters in the right panel of Fig.~\ref{fig:sur_sca}, Eq.~\eqref{eq:sur} predicts that superradiance occurs when $q > q_c = 2.03$. As shown in the plot, in both the $\sim 135^\circ < \phi < 270^\circ$ and $-90^\circ < \phi < \sim 45^\circ$ ranges, the scattered flux intensity still tends to decrease for the three values $q=1.8$, $q=1.9$, and $q=2$, whereas it begins to increase for $q=2.1$ and $q=2.2$. This trend is fully consistent with our expectation.

\section{Conclusions}\label{sec:conclusion}

We investigated the scattering cross section of massive charged scalar waves propagating in the equatorial plane of the KN BH. The effect of BH charge on the equatorial cross section is analyzed for the first time, in both the cases that the electromagnetic interaction is present and absent. In order to better present and understand the effects of BH and field parameters, in Sec. \ref{sec:numerical results} we categorize our investigation into three cases: A. the neutral scalar wave scattered by a KN BH; B. the charged scalar wave scattered by a slowly rotating KN BH; and C. the charged scalar wave scattered by a rapidly rotating KN BH. In comparison with previous studies on the scattering by a charged static BH \cite{Li:2024xyu} and the on-axis incidence on a charged rotating BH \cite{Li:2025yoz}, our study leads to the following new discoveries.

In the case A., we found that the frame-dragging effect may cause the interference minima in the equatorial cross section to appear in the exact backward direction ($\phi\approx 270^\circ$ or equivalently $90^\circ$). For a rapidly rotating BH, the equatorial cross section gradually shows an irregular oscillation as $\phi$ varies, particularly in the  $90^\circ<\phi<270^\circ$ range. In addition, we observed that the width of the interference fringes increases with increasing BH charge in the slowly rotating KN BH background, which is consistent with the cases of scattering in the RN BH and on-axis incidence to KN BH. In the case B., we found that the average intensity of the scattered flux at medium to large $\phi$ angles exhibits considerable enhancement with increasing Lorentz attraction or field mass, where the extent of the enhancement depends on the wave frequency. For the case C., due to the deformation induced by strong frame-dragging, the cross section enhancement due to stronger Lorentz attraction is not as significant. However, even in this case, the scattered flux intensity still increases with increasing field mass for waves of fixed frequency.

In Sec. \ref{sec:superradiance}, we discussed how superradiance affects the scattering cross sections. Our findings showed that the scattered flux intensity tends to increase due to superradiance within the $\sim135^\circ < \phi < 270^\circ$ range as the rotational parameter $a$ increases. This behavior can be explained by noting that the dominant contribution to the cross section in this angular range comes from the prograde partial waves with $m>0$, which are amplified as $a$ increases.  When the product of the BH and field charges satisfies the superradiant condition, we observed that the scattered flux intensity increases irrespective of the scattering direction, i.e., in both the $\sim135^\circ<\phi<270^\circ$ and $-90^\circ<\phi<\sim45^\circ$ ranges.

\section*{Acknowledgements}

This work has been supported by the National Natural Science Foundation of China.

\end{document}